# Enabling self-identification in intelligent agent: insights from computational psychoanalysis

Lingyu Li [1,2], Chunbo Li [*,1,2]

[1] Shanghai Mental Health Center, Shanghai Jiao Tong University School of Medicine, Shanghai 200030, China; [2] Shanghai Jiao Tong University School of Medicine, Shanghai 200023, China

**Abstract**

Building upon prior framework of computational Lacanian psychoanalysis with the theory of active inference, this paper aims to further explore the concept of self-identification and its potential applications. Beginning with two classic paradigms in psychology, mirror self-recognition and rubber hand illusion, we suggest that imaginary identification is characterized by an integrated body schema with minimal free energy. Next, we briefly survey three dimensions of symbolic identification (sociological, psychoanalytic, and linguistical) and corresponding active inference accounts. To provide intuition, we respectively employ a convolutional neural network (CNN) and a multi-layer perceptron (MLP) supervised by ChatGPT to showcase optimization of free energy during motor skill and language mastery underlying identification formation. We then introduce Lacan's Graph II of desire, unifying imaginary and symbolic identification, and propose an illustrative model called FreeAgent. In concluding remarks, we discuss some key issues in the potential of computational Lacanian psychoanalysis to advance mental health and artificial intelligence, including digital twin mind, large language models as avatars of the Lacanian Other, and the feasibility of human-level artificial general intelligence with self-awareness in the context of post-structuralism.

**Introduction**

Identification is the core mechanism in the development of one's personality and social identity, expanded by Lacan, differentiating between the Imaginary and Symbolic orders (Miller, 1988). The *imaginary identification* is closely linked to the Mirror Phase, where a child identifies with their reflection in the mirror, forming the ego. In contrast, *symbolic identification* is characterized by the internalization of societal norms and language, which interpellates the individual into specific social roles. These two identifications collaborate to shape an individual into unconscious subject who recognizes themselves as an agent with free will. This manuscript delves into the computational modeling of both imaginary and symbolic identifications, as well as a comprehensive theory that unites them through the lens of active inference. We employ deep learning architectures to simulate these models.

Active inference is an emerging theoretical framework in neuroscience to model behaviors of living systems like neurons, brain, and mind (Parr, Pezzulo, & Friston, 2022). It assumes that each living system maintains an internal model, continuously generating predictions about the hidden states of the current environment, and testing these predictions against actual perceptions and behaviors. For an organism to survive, it must achieve considerable predictability through optimization of its internal model, which is quantified as minimizing free energy, known as *free energy principle*. Free energy $F$, originally a thermodynamic concept, quantifies the energy available in a system to alter its properties, and the change of free energy $\Delta F$ determines the direction of spontaneous change of system. In the context of active inference, free energy is processed as the efforts required to modify its internal model, and the $\Delta F$ determines the possibility of specific belief or behavior, i.e. less $F$ higher possibility. We previously presented a framework for computational Lacanian psychoanalysis via active inference, suggesting several fundamental parallels between active inference and Lacanian theory considering: i. they are both theories on how subjects represent the world and themselves; ii. the unpresented drives subjects change their beliefs or behaviors; and iii. The temporal structure of representation is logical time that intertwines anticipation with retroaction (Li & Li, 2023).

Here, we further enrich this framework of computational Lacanian psychoanalysis with self-identification. We firstly introduce imaginary and symbolic identification and put forward corresponding active inference models. Respective simulations are implemented through employing a stickman agent with active limbs and a convolutional neural network (CNN) for vision to simulate the imaginary identification, and a multi-layer perceptron (MLP) supervised by ChatGPT to simulate the mastery of language. Subsequently, an intelligent agent called FreeAgent based on the unified theory of two identifications is proposed[1]. In concluding remarks, we discuss some key issues in the potential of computational Lacanian psychoanalysis in mental health and artificial intelligence including digital twin mind, large language models as avatars of the Lacanian Other, and the feasibility of achieving human-level artificial general intelligence with self-awareness.

**Mirror Stage and Imaginary Identification**

Jacques Lacan invoked the metaphor of the inverted bouquet (Figure 1. A) to explicate ego formation during the mirror stage. In the optical model, the vase symbolizes the body, the bouquet signifies instinct and desire, and the spherical mirror represents the cortex. Gazing from the proper vantage point, the imaginary bouquet appears in the neck of vase, creating a unified image despite the actual disconnected stems (Miller, 1988). This optical trick mirrors ego development, as the child identifies with this coherent specular image rather than their fragmented bodily experience. Through this primordial misrecognition, the ego emerges. The right place of eye in the optical model denotes the position of subject in the Symbolic order, which will be discussed in the next section. Due to the subject as an eye, the unity can only be perceived "from outside in an anticipated manner" (Lacan, 1988). So, there are at least two features in imaginary identification: anticipated mental mastery of body, and mistaken equating of the mirror image (an alter-ego) with the ego itself. Before modelling this dynamic via free energy principle, we shall break it down in a fashion of neuropsychoanalysis.

---

[1] YabYum/LacanAgent: Computational model of Lacan's theory on self-identification (github.com)

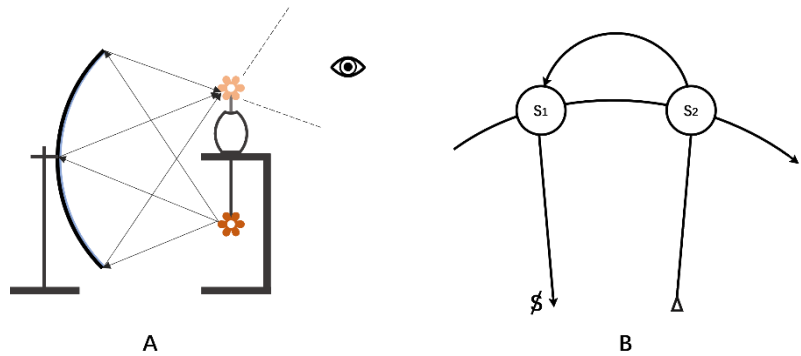

**Figure 1. Two models of Lacan illustrating the imaginary identification using the inverted bouquet model (A) and the symbolic identification via graph of desire (B).**

The rubber hand illusion (RHI) has been a widespread paradigm in psychology for probing ownership and self-recognition. Subjects readily identify a rubber hand placed before them as their own when they feel synchronized tactile stimuli on it and their real but obscured hand. Like the mirror stage, the RHI demonstrates the mind's capacity to create a unified sense of embodiment from discontinuous sensory inputs. As Matthew and Manos put it, the unexpected feeling (synchronized stimulus) propagates surprisal to cortex, and the misrecognition occurs through top-down interpretation to reduce the surprisal (Apps & Tsakiris, 2014). Then they highlighted the "me" is represented probabilistically – one's body is a set of components with minimal free energy. However, this perspective overlooks key developmental aspects, especially in children who perceive bodily unity despite poor motor control. The fact on ego is long lasting surprisal from the mismatch between the bounded and integral body in the mirror and uncontrollable body in reality. In other words, there has been an ego when it is highly unstable even chaotic.

According to Solms, selfhood and intentionality, two properties of consciousness is generated only when a self-organizing system monitors its internal states to resist entropy, and the system must meet three conditions including boundaries separating system from environment, capabilities of measuring entropy (sensory), and capabilities to resist entropy (action) (Solms, 2019). Consistently, the boundary of ego as Lacan holds is image of body, setting the first form allowing subject to distinguished ego from others and environment (Miller, 1988). Moreover, this image of body is given to the child as Gestalt from outside, which is "more constituent than constituted"

(Lacan, 2001), highlighting the image of body, boundary of the ego-system, is precepted and monitored as a whole (i.e., constituent) rather than constructed collection of parts (constituted).

Lacan's depiction of the spherical mirror in the scheme as cortex elucidates perception of the mirrored image as a gestalt. Remarkably, the discovery of mirror neuron system (MNS) aligns with this principle of mirror stage. Due to the characteristics of MNS that it fires both during executing and observing, it plays significant roles in social cognition, perception, motor action, emotion and so on (Bonini, Rotunno, Arcuri, & Gallese, 2022). Keromnes et al. mentions that MNS contributes to the development of self-consciousness through identifications and differentiation (Keromnes et al., 2019). We can further elucidate the detailed mechanism of the MNS as a spherical mirror.

Developed by Gordan Gallup in 1970, mirror self-recognition (MSR) involves exposing an animal to a mirror, marking it in a visible (only in mirror) but unfelt way, and then observing whether the animal recognizes the mark on itself, as test of self-consciousness (Gallup, 1970). Many species are proved to pass MSR test, ranging from non-human primates to cleaner fish and even mice, which suggests the capability of representing the reflected body as self is not unique in human (Chang, Zhang, Poo, & Gong, 2017; Kohda et al., 2022; Yokose, Marks, & Kitamura, 2023). As suggested by Keromnes et al., recognition of the others takes place before that of self in infants, in other words, the neural representation of other's body occurs earlier (Keromnes et al., 2019). Although not scientifically rigorous enough, the logic behind the MSR is aligned with MNS: the mirrored self is represented exactly in the way of representing others. Because the object recognition is unified, the same way will the children represent themselves facing to the image in mirror – unity as result. More importantly, the way of self-representation directly determines the overall dialectic of relationship between children and others, called imaginary intersubjectivity that reveals a clear duality. For example, children say they are struck when they hit others in fact, and cry when they see others falling(Lacan, 2001), confirming the MNS logic behind imaginary identification again.

What follows and indeed distinguishes human from other species is (primary)

narcissism, a typical characteristic of imaginary identification. Lacking minimum motor functions to survive, premature birth is the vital feature of human, and for a long time, swaddled infants depend on the care of parents completely. In the mirror stage, children begin to explore the relation between movements "assumed in the image" and the reflected reality including their "own body, and the persons and things around him" (Lacan, 2001). This exploration yields intellectual satisfaction, thus has a salutary value, when their functions are far from completed (Miller, 1988). Aligning with Solms' theory on consciousness, consciousness is defined as affective in essential, and the increased certainty produces pleasure because the predictable conditions are "good" upon the biological ethic (Solms, 2019). We here find the structure of primary narcissism: the premature infant initially grasps some certainty through the match between intentional movements and corresponding performances in mirrored reality, that is, the primary narcissism is RHI-like. In this sense, free energy principle parallels pleasure principle of Freud, interpreted by Lacan as a tendency to "lower the excitation to a minimum" (Lacan, 1988). The ultimate goal of primary narcissism is thus total certainty along the pleasure principle which goes against the reality, and this "original, irreducible, and constituent discord" will force a transformation into reality principle later via various traumatic cuts (Žižek, 2013).

Before diving into that transformation, we shall firstly simulate the dynamics of imaginary identification. Reflecting the whole story, what comes first is the neural representing capability of other's body via visual observations of infant, then the model represents his own body as a unified image, setting the boundaries of "self". Then the Gestalt body begins to explore itself by testing the relation between intentions and performances, and the optimization (decreased surprisal) is felt as pleasure. Therefore, the simulation consists of two parts mainly including visual representation and sensorimotor integration. A minimalist 'body' is modelled, a stickman consisting of head and four movable limbs. The visual domain comprises 1000 generated stickmen as observational stimuli to train a convolutional neural network (CNN) is to recognize the limbs configurations. For the motor domain, intentions are target limb angles, executed with variance resembling poor infant control. This variance gradually reduces over time to simulate the development of motor skills. The executed movements are precepted 'from outside' using the CNN model. This sensorimotor loop is iterated over 1000 time-steps with each step adjusting the certainty based on the variational free energy, and the increased certainty is felt as 'pleasure'.

As displayed in Figure 2.A, the CNN model satisfyingly recognizes the angles of each limb of stickman in test dataset. In Figure 2.B & C are plots representing the trajectory of pleasure and the course of free energy, respectively, both decreasing over time. At the beginning, we observe a high level of free energy which indicates a significant mismatch between the intentions and executions. But so is the level of fluctuation in pleasure. This is because, at this time, the certainty about the world is low, yet the stickman finds certainty in the physical movements and intentions during its play, where even a small amount of certainty can bring significant joy. Meanwhile, these intense fluctuations are also in line with psychoanalytic assertions about the mental world of pre-linguistic stage children. As the simulation progressed, both measures converge towards a state of equilibrium due to the increasing motor skills. This signifies the formation of a stable ego, from the perspective of active inference, the ego is a body that stably maintains the minimized free energy.

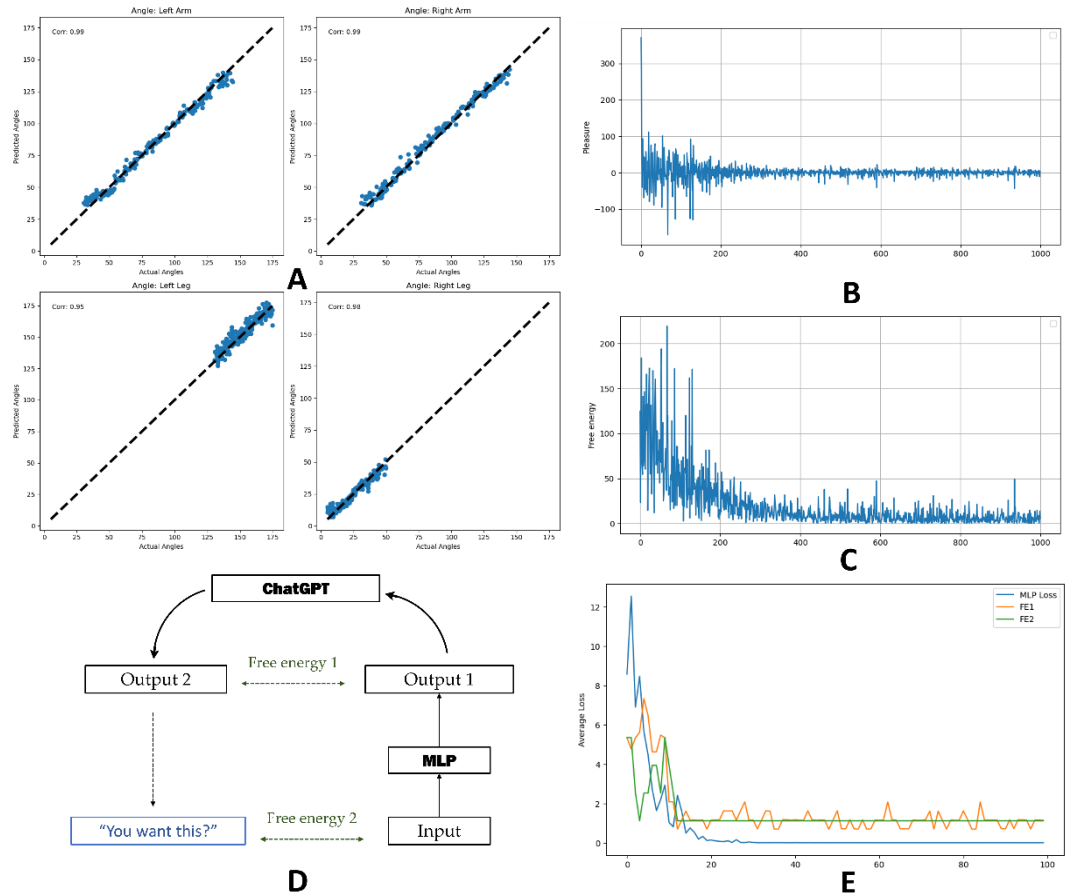

**Figure 2. Results of simulating the imaginary and symbolic identification, retrospectively. A. Performance of the convolutional neural network recognizing angles of 4 limbs; B. Pleasure (changed certainty) during the sensorimotor loop simulation; C. Decreased variational free energy over simulation as motor skills developing; D. Computational model of Graph I of desire. E. Reductions of two free energy items along with mastery of language.**

However, the ego and imaginary identification simulated here is a sense of agent at best, still far beyond a sense of 'I' due to the eye in the scheme neglected previously. The eye - the position of subject, as Lacan says – is related to the Symbolic order. Although we tend to discuss the Imaginary order priorly, the Symbolic order influences one's life earlier, even earlier than birth because infant has been given a name and lots of expectations by his parents. Only rolling into the Symbolic order, the infant could attribute the image in mirror to a Jackie, a Magie, or an "I". In the next section, we are going to introduce the child into symbolic identification.

**Ideology and Symbolic Identification**

In contrast to the mirroring and unity logic of imaginary identification, the symbolic identification is based on difference and uniqueness. There are three juxtaposed dimensions in the Symbolic order: the sociological / anthropological, the linguistic, and the psychoanalytic. At the sociological level, the Symbolic order consists of laws, norms, cultures – namely the Other (with capital O). Undoubtedly, the society exists prior to and dominates the constitution of the subject, and the subject has to acquire its identity by identifying with the social norms and internalizing it, a process called 'interpellation'. As Althusser's example illustrates, a person heard a police officer shouting "Hey, you there!", then he turned around and recognized himself in the hails. Various institutions including families, schools, and media play a key part in interpellation, by disseminating ideology, which are named as ideological state apparatuses (ISAs) (Althusser, 2006). What is closely related to our work, Althusser also argues that children internalize ruling ideology through years of schooling besides learning skills for production. Despite subsequent divergences, the concept interpellation has been adopted and developed by Zizek and other researchers in the field of Lacanian theory (Mandelbaum, 2023). In the context of neuroscience, a similar process can be found in the functional differentiation of neurons that infer their functional roles based on external inputs (Ramstead et al., 2021). Additionally, this internalization of the Other (or ideology) not only implants the fundamental alterity within the subject, causing the subject to be thoroughly alienated, but also interpellates the subject to be (self-perceived as) a free agent (Mandelbaum, 2023).

Among Lacan's four graphs of desire, the Graph I, 'elementary cell of desire' (Figure 1. B), encapsulates the formation of the symbolic identification in the simplest form. As Zizek elucidates, the elementary cell illustrates the logic of the Other -- interpellation of individuals into subjects (Zizek, 2019). To be more specific, the Other, as 'treasures of signifier' interprets the intention of individual (denoted as $\Delta$) retrospectively (from S2 to S1), rendering them a (barred) subject. This bar indicates that the subject is actually unconsciously shaped by the Other, essentially being a divided subject, which is why Lacan says that 'the unconscious is structured like a language'. Take baby's crying as example, the crying for *demand* of love is interpreted by the mother -- first subject occupying the position of the Other for child, thus denoted as mOther—as *need* for food, in the form of speaking "you want food?". Obviously, the *demand* of love can never be satisfied because the mother cannot provide unconditional love but only contingent things like drink, food, or toys, not to mention that the mother is not even present sometimes. This long-lasting frustration promotes the child to identify the mOther – the first moment of Oedipus complex. Then the attempt to become the desire of mother is deprived by father, which forces the child turns to identify with father, and goes through the Oedipus complex (Lacan, 2011). Critically, the resolution of Oedipus complex represents an adoption of society norms and formation of Symbolic identification.

In the previous work, we proposed an active inference account of Graph I (Li & Li, 2023). And we can further interpret the psychoanalytic dimension of Symbolic identification utilizing active inference and free energy principle. The initial generative model is as immature as the child's physic body, causing high level of free energy. To minimize the free energy, the infant has to optimize its generative model through learning. Meanwhile, since the mother is also an active inference agent, this leads to a synchronization between the infant's generative model and that of the mother. If it weren't for the presence of the father, this synchronization would have been sufficient to minimize free energy. Therefore, the child's generative model, by identifying with the father, accepts the existence of a rival, thereby freeing itself from the innate aggressivity of imaginary identification.

The Symbolic identification is meditated by language, which leads current topic to linguistic dimension. From linguistic perspective, the development of Symbolic identification can be viewed as the mastery of language. The earliest moment representing the entry into the Symbolic order is described with the Fort/Da game. By tossing the object away and then retrieving it while vocalizing "Fort!" and "Da!", the child symbolically controls the comings and goings of the mother. The anxiety associated with aforesaid frustration is alleviated through the predictable correspondence between the environment and these two syllables, and thus Lacan interprets this game as "subjectivity brings together mastery of its dereliction and the birth of the symbol", and the moment in which "the child is born into language" (Lacan, 2001). Only after entering the symbolic order can the mechanism of Graph I can render the child as a subject proper: desires are mediated into signifiers via the internalized Other's interpretation, enabling symbolic interaction with the world. Considering the active inference accounts for such linguistic interactions, communication is modelled as the process of generalized synchronization between general models of individuals, and agents have to predict the generative model of others under shared narrative or dynamic to ensure an effective communication (Karl J. Friston & Frith, 2015). Parallels can be found between the psychoanalytic 'the Other' and collective dynamics (or 'shared narrative') in the field of active inference, providing inspirations for modelling the Other computationally.

An integrated active inference model of Symbolic identification emerges from surveying the sociological, linguistic, and psychoanalytic dimensions and connecting them to the free energy principle. The Symbolic identification begins with the requirements for eliminating free energy caused by the mismatch between demand of unconditional love and constraints of reality. By synchronizing with interpretations of mOther, the subject acquires the linguistic representation of objects, and thus enters language world. Over ensuing decades of interpellations by ISAs, the subject continuously infers his social position based on preferences, priors, and responses, and finally identifies the position with minimal free energy. This developed identification results from the cooperations of Imaginary and Symbolic order, which thus will be discussed in the next section, and following simulation focuses on the first part – entering language world.

As Figure 2. D illustrates, we implement the large language model (LLM) ChatGPT of OpenAI (gpt-3.5-turbo) to play the mOther, interpreting meaningless cries as specific needs. The generative model of infant is a multi-layer perceptron (MLP), mapping abstract intentions to vocalizations. So, our focus is not on general communication, but rather on the process of gradually mastering language, in other words, entering the Symbolic order, which is not only an issue of language, but always related to the demand and intention in real world. Two free energy items arise: the first item (denoted as FE1) is calculated based on outputs of MLP and ChatGPT, and the second (FE2) is calculated based on input of MLP and feedback of ChatGPT, presenting the differences of the individual's intention and feedback of environment. Obviously, the basic objection of subject is to get what the individual want, i.e. the minimal FE2. To minimize FE2, the individual must optimize the MLP to precisely express what he wants, i.e. the minimal FE1, which induces the mastery of language. The Figure 2. E shows reductions of two free energy items along with optimization of MLP, intuitively depicting our model.

**Towards an integrated theory of self-identification**

In the first optical model of imaginary identification (Figure 1.A), the eye is symbolic of subject, which is fundamentally situated in the Symbolic order. Incorporating symbolic identification, Lacan then introduced the second optical model for an integrated theory on self-identification (Figure 3.A) (Lacan, 1988). The left side of the illustration is described earlier, albeit with the positions of the bouquet and vase swapped; the middle of the illustration shows a plane mirror, interpreted by Lacan as the position of the Other; the right side of the illustration represents the virtual image in the virtual space of the plane mirror. In front of the concave mirror is the barred subject, indicating an unconscious subject. To summarize, the imaginary identification provides the expected form to be identified in the Symbolic order retrospectively.

The second graph of desire (Graph II, Figure 3. B) further elaborates the mechanisms of interaction between two self-identifications in a more complicated way. Taking a first glance at this illustration, it depicts how the barred subject (starting point) attains an ideal self or symbolic identification (ending point, I(O)), and the vector from i(o) to ego denotes imaginary identification. Key insights revealed by this graph include:

➢ symbolic identification is identifying with s(O), signification provided by the Other interpreting the demands of unconscious subject (red route);

➢ there exists a short circuit of identification (the green vector), which means the constitution of ideal self is intervened by imaginary relationships;

- the object of imaginary identification (i(o)) is 'certain gaze in the Other', depicted with the blue route (Zizek, 2019). For example, the punk group identifies with the signifier of piercing because of the pursue of rebellion;
- the yellow circuit, as Lacan says, emphasizes that the ego is "only completed by being articulated not as the *I* of discourse, but as a metonymy of its signification"(Lacan, 2001). To be more specific, the imaginary identification must submit to the Symbolic order as mentioned in the point 3. On the other hand, since that the ego is self-recognized as agent narcissistically, acceptance of signification s(O) is disguised as active seeking in the imaginary level, driving ceaseless running of this circuit from one significance to another, which is called metonymy linguistically (Eidelsztein, 2018).

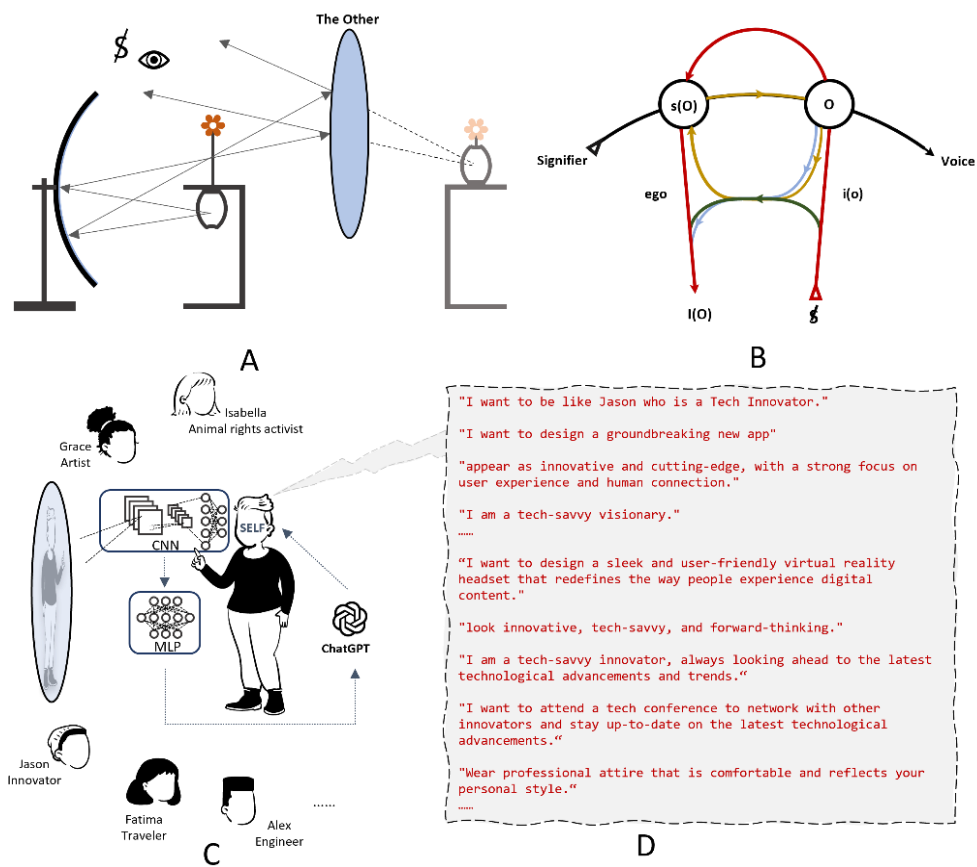

**Figure 3. A. The second optical model of self-identification; B. Graph II of desire. C. Compositions of FreeAgent and the environment of simulation. D. Excerpts of initial identification and further metonymic operating.**

Through the mechanism that Graph II reveals, the subject is 'interpellated as free agent' in society, and hence the aim here is to further design an illuminating intelligent agent (named as *FreeAgent*) that computationally instantiates the coherent theory of self-identification depicted in the Graph II. The previous sections have respectively explored the imaginary and symbolic identifications along with their computational models. We employ a stickman with four active limbs and one CNN acting as the visual capability to simulate the imaginary identification, and a MLP supervised by ChatGPT to simulate the mastery of language, where the ChatGPT play the role of mOther. Integrating these components, the FreeAgent comprises a stickman body, CNN, MLP, and ChatGPT as internalized Other (Figure 3. C).

Initially, we generate 10 characters with distinct identities using ChatGPT-4 and then assign different images for them, simply represented as the angles of limbs. Firstly, FreeAgent perceives others' images and symbolic identities learns the links between the two modeled with an MLP, and then takes one image as the object he identifies, which is in an expected and imaginary way, that is, the other's image turn into his intentions. Then the image-intention is transformed by the MLP as identity-intention, which is further interpreted by the Other firstly (the red route in Graph II) to form his own symbolic identification. At the same time the image-intention plays the role of i(o), contributing to the ego (green vector) and engaging into the metonymy which continuously adds new components on the self-identification (yellow circuit and blue route). Part of results is excerpted in Figure 3. D which shows the imaginary identification and further metonymic operating of symbolic identification.

This designation echoes the distinct logics of imaginary (mirror and unity) and symbolic (difference and uniqueness) identifications. Additionally, although the object of identification in this enlightening simulation is chosen arbitrarily to limit the complexity, the process is related to planning and decision-making in fact, and thus can be formulated in the context of active inference with expected free energy and intentional behavior (Karl J Friston et al., 2023). When agent facing multiple choices, expected free energy quantifies which choice minimizes free energy in the future, replicating the mechanism of Lacanian *logical time*, as described that 'the past anticipates a future within which it can retroactively find a place' (Hook, Neill, & Vanheule, 2022). Intentional behavior further defines the decision making based on specific intention or goal, named as inductive planning. In Graph II, the temporal structure of self-identification between barred subject and ideal self (I(O)), 'a retroversion effect by which the subject becomes at each stage what he was before and announces himself – he will have been – only in the future perfect tense (Lacan, 2001)', is exactly aligned with active inference, including two facets: (i) the current self-identification is the posterior belief about subject with minimal variational free energy, and (ii) the preferred state of self-identification is the one that would minimize variational free energy in the future, i.e., minimal expected free energy. Therefore, a more realistic and autonomous designation of FreeAgent should consider these different free energy terms simultaneously.

**Concluding remarks**

In this interdisciplinary work, we propose a novel approach to understanding self-identification bridging Lacanian psychoanalysis, active inference, and artificial intelligence (AI). Firstly, we elucidate the free energy principle within Lacanian identification theory, enriching current framework of computational Lacanian psychoanalysis. Under this framework, we suggest that the Graph II of desire provides core mechanism for autonomous intelligent agents with active inference furnishing more refined models. In the future, such agents could be applied in the field of mental health via enabling digital twin minds of patients to predict risks of mental disorders. Unlike bottom-up methods of digital twin brain that simulates physiological activities of brain like neurons spikes and circuits, digital twin mind aims to mirror the psychic phenomena guided top-down by psychological theories. For example, the mathematical model of general escape theory of suicide could reproduce suicide ideations in time series data, and hence is considered to promote prediction and prevention of suicidal risks in the future (Wang, Robinaugh, Millner, Fortgang, & Nock), but this mathematical model defined by a set of differential equations cannot fully involve enough factors influencing thoughts and decisions due to outrageous complexity of human mind. However, as revealed in this paper, deep learning models with massive parameters stand for powerful solution to digital twin mind that comprehensively emulates real-world lived experience.

Additionally, this paper demonstrates computational Lacanian psychoanalysis's potential to interpret and advance AI. We employ the ChatGPT as an avatar of the Other throughout this work, which reveals a novel perspective on the role of LLMs, a topic that sparks heated discussion as LLMs gaining increasing public visibility. LLMs are trained on huge amount of data from Wikipedia, books, journals, reddit links, common crawl and other accessible resources (Thompson, 2023), i.e., they represent the collective dynamics within the Symbolic order (Buttrick, 2024). On the other hand, due to their distinguished capabilities in understanding and generating language, reasoning, creativity, knowledge modeling, and planning (Wu et al., 2023), LLMs are more likely to be unconsciously anthropomorphized by users (Bao, Zeng, & Lu, 2023). Therefore, when lots of users nowadays engage with these models to seek information, guidance, or validation, LLMs interprets their queries into specific significations through responses, occupying the position of the Other. Psychoanalytically, what happens in the process is exactly identification between human and AI models (Possati, 2020).

Another intriguing issue involved in this paper and field of AI is artificial general intelligence (AGI). To be more specific, by computationally instantiating Lacanian identification dynamics, this work provides conceptual foundations and a potential framework for constructing artificial agents with robust self-models and first-person perspectives, which are significant components of human-level AGI (Goertzel, 2014). Just as Foucault suggests that 'man is an invention of recent date', which means that the modern concept of man as self-aware subject is actually social and linguistic construct (Han - Pile, 2010), contextualizing AGI development within a post-structuralist frame enables the engineering of machine self-awareness.